# Fe alloy slurry and a compacting cumulate pile across Earth's inner-core boundary

Youjun Zhang[1,2*], Peter Nelson[3], Nick Dygert[3,4], & Jung-Fu Lin[3,*]

[1]Institute of Atomic and Molecular Physics, Sichuan University, Chengdu 610065, China.

[2]Center for High Pressure Science and Technology Advanced Research (HPSTAR), Shanghai 201900, China.

[3]Department of Geological Sciences, Jackson School of Geosciences, The University of Texas at Austin, Austin, TX 78712, USA

[4]Planetary Geoscience Institute, Department of Earth and Planetary Sciences, University of Tennessee, Knoxville, USA.

# Fe alloy slurry and a compacting cumulate pile across Earth's inner-core boundary

**Youjun Zhang [1,2], Peter Nelson [3], Nick Dygert [3,4], & Jung-Fu Lin [3,*]**

[1]Institute of Atomic and Molecular Physics, Sichuan University, Chengdu 610065, China.
[2]Center for High Pressure Science and Technology Advanced Research (HPSTAR), Shanghai 201900, China.
[3]Department of Geological Sciences, Jackson School of Geosciences, The University of Texas at Austin, Austin, TX 78712, USA.
[4]Planetary Geosciences Institute, Department of Earth and Planetary Sciences, University of Tennessee, Knoxville, Knoxville, TN 37996, USA.

*Corresponding authors. E-mail address: afu@jsg.utexas.edu (J.-F. Lin) or zhangyoujun@scu.edu.cn (Y. Zhang)

## Abstract

Seismic observations show a reduced compressional-wave gradient at the base of the outer core relative to the preliminary reference Earth model and seismic wave asymmetry between the east-west hemispheres at the top of the inner core. Here, we propose a model for the inner core boundary (ICB), where a slurry layer forms through fractional crystallization of an Fe alloy at the base of the outer core (F layer) above a compacting cumulate pile at the top of the inner core (F' layer). Using recent mineral physics data, we show that fractional crystallization of an Fe alloy (*e.g.*, Fe-Si-O) with light element partitioning can explain the observed reduced velocity gradient in the F layer, in cases with a solid fraction of ~15±5% in liquid with a compositional gradient due to preferential light element partitioning into liquid. The compacting cumulate pile in the F' layer may exhibit lateral variations in thickness between the east-west hemispheres due to lateral variations of large-scale heat flow in the outer core, which may explain the east-west asymmetry observed in the seismic velocity. Our interpretations suggest that the inner core with solid Fe alloy has a high shear viscosity of ~$10^{23}$ Pa·s.

**Main text**

**1. Introduction**

Our knowledge of the Earth's core is mainly derived from seismic observations and mineral physics studies, which show the core comprises of a solid inner core surrounded by a liquid outer core. The inner core grows as the molten outer core cools and solidifies, releasing energy that drives the geodynamo and generates Earth's magnetic field (Buffett, 2000; Davies et al., 2015). Recently, detailed studies of PKP, PKIKP and PKiKP differential travel times and amplitudes show a reduced P-wave velocity ($V_p$) gradient layer compared to the PREM at the bottom ~280 km of the outer core, known as the F layer (Fig. 1) (Adam et al., 2018; Dziewonski and Anderson, 1981; Kennett et al., 1995; Song and Helmberger, 1995; Souriau and Poupinet, 1991; Zou et al., 2008). Most seismic observations indicate that the F layer is global, surrounding the entire inner core (Cormier et al., 2011; Souriau and Poupinet, 1991; Zou et al., 2008). Additionally, PKIKP and PKiKP waves show a hemispherical dichotomy at the top ~200 km of the inner core which we call the F' layer (Fig. 1) (Deuss et al., 2010; Monnereau et al., 2010; Niu and Wen, 2001; Waszek and Deuss, 2011; Yu and Wen, 2006). In the F' layer, the eastern hemisphere has a faster $V_p$ than the western hemisphere in Fig. 1 (Monnereau et al., 2010). The upper ~100 km of the F' layer has an isotropic velocity variation of ~1.5% between the two hemispheres, while in the deeper part the difference decreases (~0.5%) (Deuss, 2014). Understanding the cause of the reduced $V_p$ gradient in the F layer and hemispheric asymmetry in the F' layer is crucial to reveal the inner core's formation and dynamics, in addition to the thermal and compositional state of the inner and outer core.

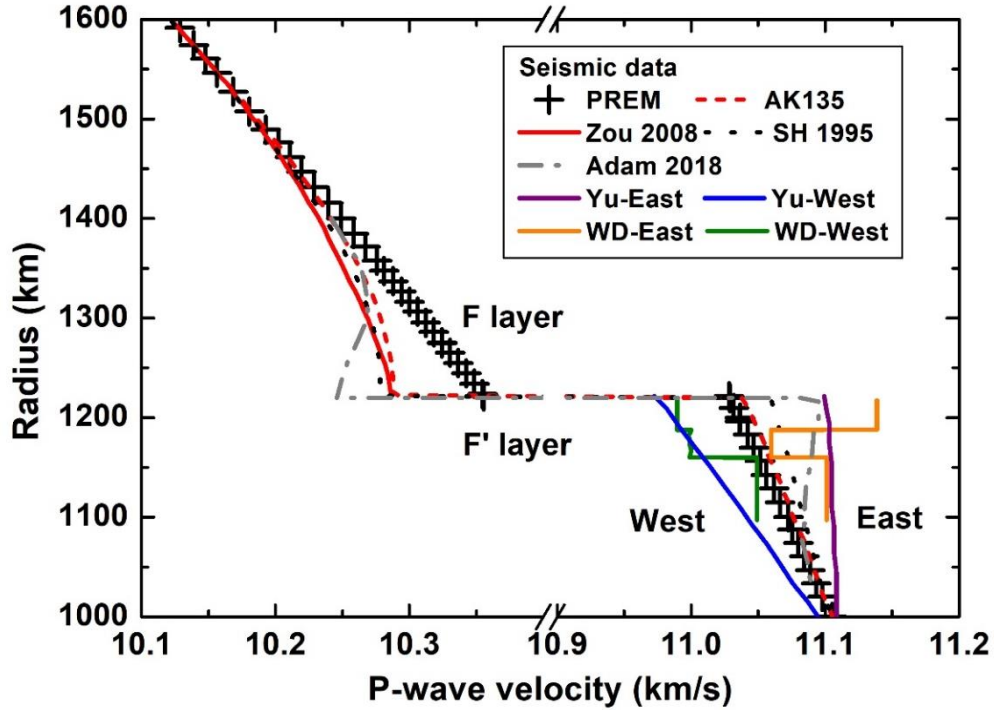

**Fig. 1. P-wave velocity across the inner-core boundary.** Data from seismological observations are shown for the PREM (plus symbol line) (Dziewonski and Anderson, 1981), AK135 (red-dashed line) (Kennett et al., 1995), Zou 2008 (red-solid line) (Zou et al., 2008), SH 1995 (black-dotted line) (Song and Helmberger, 1995), Adam 2018 (gray dash dot line) (Adam et al., 2018), Yu-East (purple line) (Yu and Wen, 2006) and Yu-West (blue line) (Yu and Wen, 2006), WD-East (orange line) (Waszek and Deuss, 2011) and WD-West (olive line) (Waszek and Deuss, 2011), respectively.

Investigations of seismic phenomena at the ICB link them to Fe alloy solidification and the formation and evolution mechanisms of the inner core (Deguen, 2012). With the recently refined seismic observations across the inner-core boundary (ICB, Fig. 1) (Yu and Wen, 2006; Zou et al., 2008), some proposed models have tried to reconcile seismic observations with more reasonable geophysics and geodynamics for the F layer and F' layer (Deuss, 2014). Particularly, solidification or melting of Fe with light element partitioning has been suggested to cause light-element-rich or light-element-poor variations in the F layer. For example, Gubbins et al. (2008) proposed that $V_p$ variations in the F layer can be explained by a stably stratified layer on the liquidus with compositional variations with depth due to solidification and remelting of Fe alloy (Gubbins et al., 2008). The model explains that the reduction in seismic velocity with depth is caused by light element(s) being released from the F layer. A thermochemical flow model was also proposed by Gubbins et al. (2011), arguing that variations in mantle heat flow may result in temperature variations and localized melting and freezing at the ICB due to large-scale convection patterns in the outer core. Subsequently, lateral variation at the top of the inner core such as the hemispheric asymmetry in the F' layer (Gubbins et al., 2011) may be generated. However, this model does not use mineral physics data to explain seismic observations such as the origin of the east-west hemispherical asymmetry, nor explain the processes of the liquid Fe alloy crystallization (and remelting) in detail.

Another scenario invoked to explain the east-west asymmetry is an inner core translation model, where the F' layer is generated by the simultaneous solidification and melting of the western and eastern hemispheres, respectively (Aboussiere et al., 2010). This process results in a lateral translation of the inner core material from west to east, causing a lopsided growth of the inner core (Aboussiere et al., 2010; Monnereau et al., 2010). However, this inner core translation model is unlikely because it does not explain the existence of a sharp hemispheric boundary in the F' layer (Monnereau et al., 2010). Furthermore, inner core convection may not occur if thermal conductivity is as high as recently suggested (Davies et al., 2015; Ohta et al., 2016). Thus, it remains challenging to explain the origin of the reduced seismic $V_p$ gradient in the F layer and hemispheric asymmetry in the F' layer simultaneously.

Besides these above models, a slurry F layer was proposed by S.I. Braginskii in 1963, which explains the compositional convection process in the Earth's core. It was further investigated by (Fearn et al., 1981), (Loper and Roberts, 1981), (Loper, 1983), (Sumita et al., 1996), and others, from the geodynamic and thermodynamic state of the core to understand possible formation mechanisms of the slurry layer. A recent study indicates that a slurry F layer could satisfy the geophysical constraints on the density jump across ICB and the core-mantle boundary (CMB) heat flux as well (Wong et al., 2018). An approximately 100-km-thick slurry layer likely exists at the top of the present Martian core as well, according to a geodynamical model with magnetic and geodetic constraints (Davies and Pommier, 2018). Therefore, a universal core formation process may exist for terrestrial planets, where solid cores grow from the sedimentation of suspended particles snowing down from a slurry zone above their solid core boundary (Breuer et al., 2015; Buffett et al., 2000; Stewart et al., 2007; Sumita et al., 1996). A compacting pile or mushy zone may form near the top of the solid boundary as a result of the particle sedimentation in the slurry (Tian and Wen, 2017), which is partially molten, analogous to a cumulate pile at the bottom of a magma chamber (McKenzie, 2011; Shirley, 1986; Sumita et al., 1996). The aforementioned slurry layer models have yet to be integrated with recent mineral physics results, or to address the mechanism of the slurry formation and the consequent inner core freezing.

In this study, we develop a geodynamic model that includes fractional solidification and light-element segregation in an Fe alloy system, and invoke sedimentation of solid particles across the inner-outer core boundary. We infer that a compacting cumulate pile forms beneath the ICB (the F' layer). We examine the phase diagram, sound velocity-density profiles, and viscosity of an Fe alloy core across the ICB to explore the implications of the dynamical model. This scenario is consistent with mineral physics and seismological constraints at the ICB conditions and provides a holistic model of inner-core growth.

## 2. The composition of Earth's core

The Earth's core is known to be composed of an Fe light-element alloy to balance its density deficit compared to pure Fe-Ni. Several light element candidates have been proposed, such as Si, O, S, C, and H (Poirier, 1994a), although which of these elements is the most abundant in the alloy is still a matter of intense debate (Li and Fei, 2014). The presence of a combination of these light elements is necessary to explain the density deficits and velocity discrepancies between measured pressure-density-sound velocity profiles of Fe and the seismological observations. Some of the proposed major light elements have a very low solubility in Fe at ambient and high pressure-temperature (P-T) conditions such as C and H (Li and Fei, 2003). Additionally, C and H have high volatility, so they are unlikely to significantly incorporate into Fe at the early differentiation of the core formation (McDonough and Sun, 1995). S is a strong siderophile but moderately volatile element, whose geo- and cosmochemical constraints give an upper limit of ~2 wt% S in the core through mass balance calculations of the bulk Earth compositions (Dreibus and Palme, 1996). Therefore, Si and O are likely the most abundant light elements present in the Earth's core, as supported by recent high-pressure mineral physics, Si isotope data, and some core formation models (Georg et al., 2007; Hirose et al., 2013; Huang et al., 2011; Lin et al., 2002; Mao et al., 2012; Zhang et al., 2016).

A density jump inferred from seismic data (~0.5–0.6 g/cm$^3$) is present across the ICB (Dziewonski and Anderson, 1981; Kennett et al., 1995) (Fig. 2), which may be attributable to light-element release during crystallisation, suggesting that the inner core contains approximately 30–50% less light elements than the outer core (Anderson and Isaak, 2002; Fei et al., 2016). Some theoretical calculations have suggested that O partitions more strongly than Si from solid into liquid (Alfè et al., 2002), so the solid inner core may be O poor on freezing. Recent melting experiments of an Fe-Si-O alloy at high P-T in a DAC also found that the Fe-Si-O alloy could crystallize $SiO_2$ as it cools so that O and Si would partition from solid to liquid until one of the two is mostly depleted in the core (Hirose et al., 2017). Consequently, we adopt a compositional model for the Earth's core where Fe-Si-O exists in the liquid outer core, while Fe-Si is in the solid inner core. We cannot exclude the coexistence of other light element(s) but in this study, we use Fe-Si-O system as our compositional model because of recent experimental results.

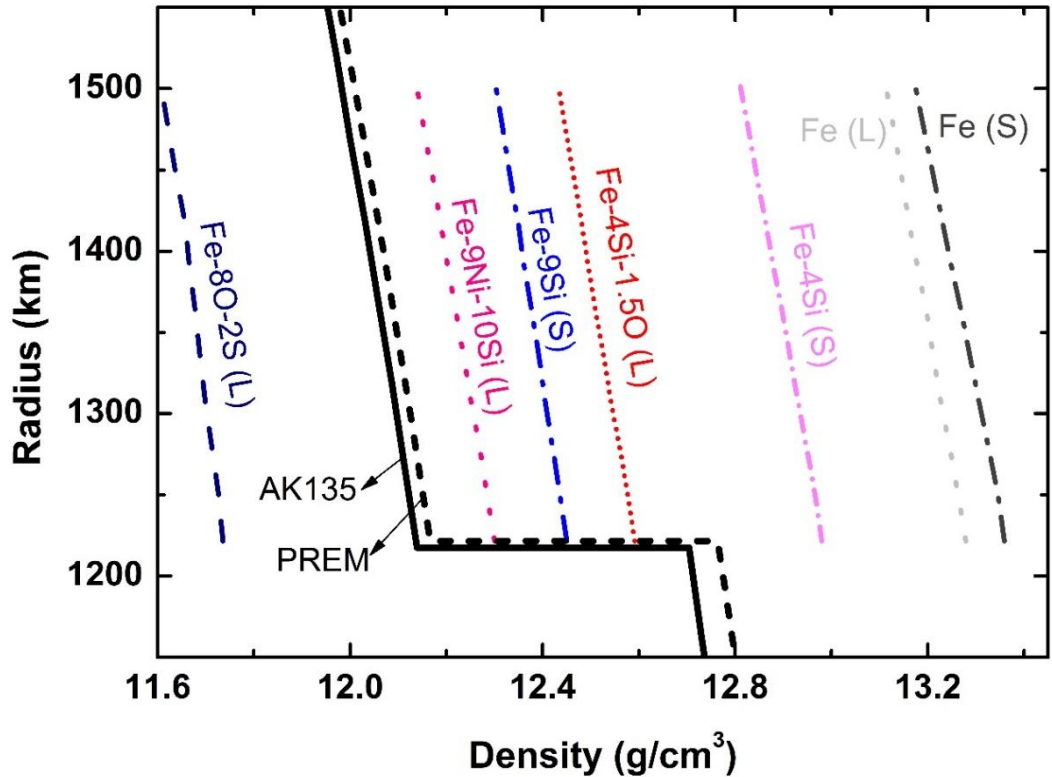

**Fig. 2. Densities of the Fe and Fe alloys in the lowermost outer core (F layer).** Densities of liquid Fe-4Si-1.5O and solid Fe-4Si are higher than those of the outer core and the inner core, respectively, but their density contrast (~0.4 g/cm$^3$) is close to the AK135 and PREM models (~0.5–0.6 g/cm$^3$) (Dziewonski and Anderson, 1981; Song and Helmberger, 1995). The symbols "L" and "S" represent the liquid and solid state, respectively. Densities at the relevant conditions of the ICB are shown for *hcp*-Fe (black dashed-dotted line) (Dewaele et al., 2006; Lin et al., 2005; Mao et al., 2012), liquid Fe (gray-dashed line) (Brown and McQueen, 1986; Ichikawa et al., 2014; Nguyen and Holmes, 2004), *hcp* Fe-9wt.%Si (blue dashed-dotted line) (Fischer et al., 2014), liquid Fe-9wt.%Ni-10wt.%Si (pink-dotted line) (Zhang et al., 2016), and liquid Fe-8wt.%O-2wt.%S (navy-dashed line) (Huang et al., 2011).

The densities and sound velocities of Fe/Fe-Si/Fe-O alloys have been well investigated at the relevant conditions of Earth's core (Huang et al., 2011; Lin et al., 2005; Mao et al., 2012; Zhang et al., 2016). Considering Si and O as the light elements in the outer core, a release of ~2 wt.% O upon freezing would explain the density contrast of ~4% between the inner and outer core (Fig. 2) (Fischer et al., 2011; Huang et al., 2011). Comparisons of the experimentally determined and modelled sound velocity of Fe and Fe-Si/Fe-O alloys with the PREM suggest a Si-rich Fe core with an outer core composition of ~4–6 wt.% Si and ~1–3 wt.% O, and an inner core of ~4–6 wt.% Si (Fig. 3) (Hirose et al., 2013; Zhang et al., 2016).

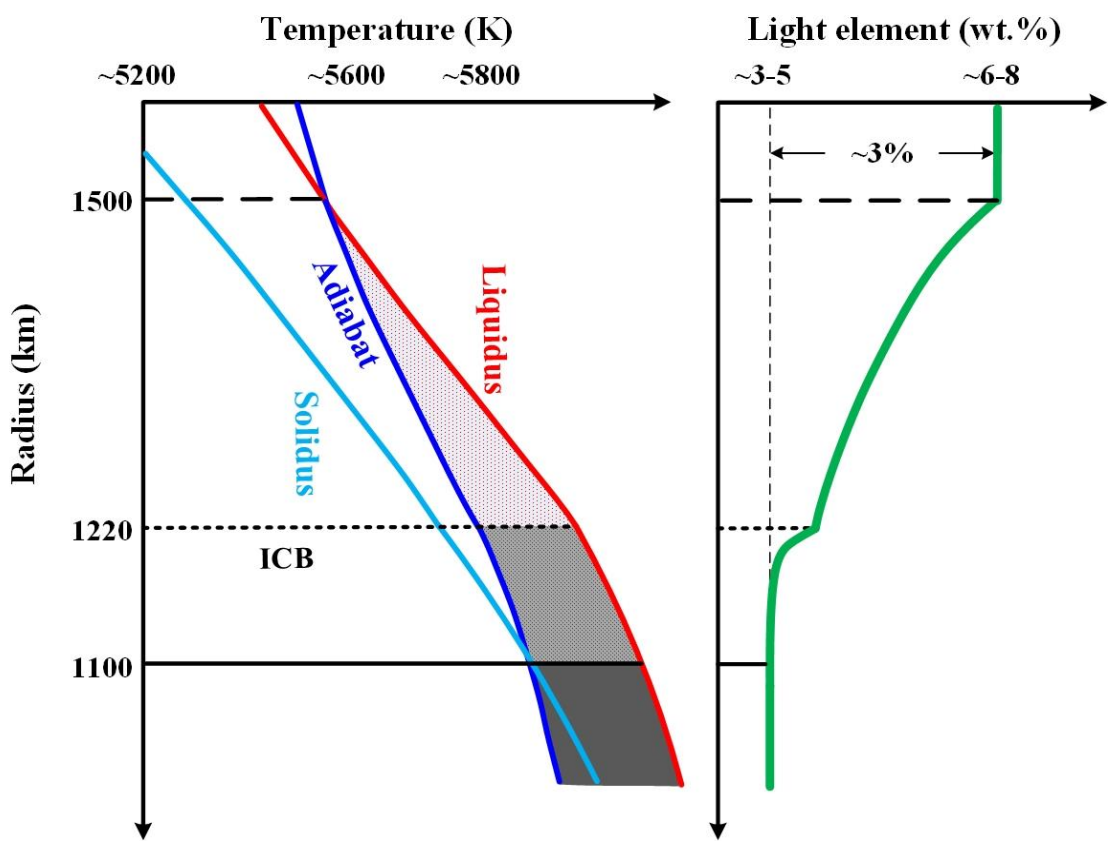

**Fig. 3. Schematic thermal structure and concentration of light elements across the inner-core boundary**. Red and light-blue lines represent the liquidus and solidus of Fe-Si-O alloy, respectively; blue line shows the adiabat of Earth's core across the ICB;

green line is the light element concentration of the core composition across the ICB; dashed lines and dotted lines represent boundaries between the outer core and F layer and between F layer and F' layer, respectively.

Since the Earth's inner core solidifies across the ICB, a mixture of liquid and solid Fe light-element alloy is expected to co-exist at the ICB conditions within the liquidus and solidus of the system (Fig. 3). Previous studies show that adding light elements such as Si, O, and S into Fe depress its melting temperature (*e.g.*, (Fischer, 2016; Morard et al., 2014). Fe-Si alloy has a higher melting temperature compared with Fe-O and Fe-S alloys so that Fe-Si would preferentially crystalize as the core cools (Huang et al., 2010; Mori et al., 2017; Zhang et al., 2018).

**3. Geodynamic model and mineral physics results**

**3.1. Thermal structure and geodynamic model across the ICB**

The cooling of Earth's core is a key factor in the crystallization process and the heat transfer in the core boundary layer. The liquid core starts to crystallize when the temperature (adiabat) reaches its liquidus (Fig. 3); with further cooling, the core grows and by conductive cooling into the overlying mantle. Pure Fe has a single melting curve at high pressures, while a mixed core composition such as Fe-O-S/Fe-Si-S system has both a solidus and liquidus with a temperature gap (Fischer, 2016; Morard et al., 2014); (Sakairi et al., 2017; Terasaki et al., 2011) (Fig. 3). In our model, the core adiabat crosses the liquidus at the top of the F layer, while it is still higher than the solidus, as shown in Fig. 3. As a result, liquid starts to crystallize at the top of the F layer and thus solid is crystallizing through the F layer.

The liquidus temperature of the Fe alloy increases with increasing depth (pressure) and with the accompanying decrease of the light element concentration. The temperature difference between the adiabat and liquidus steadily grows from the radius of 1500 km to 1220 km, allowing for continuous solidification (Fig. 3). Although the process we proposed is not strictly adiabatic, consideration of a hypothetical core adiabat and its relationship with the core liquidus allows us to evaluate the potential for the formation of a slurry layer.

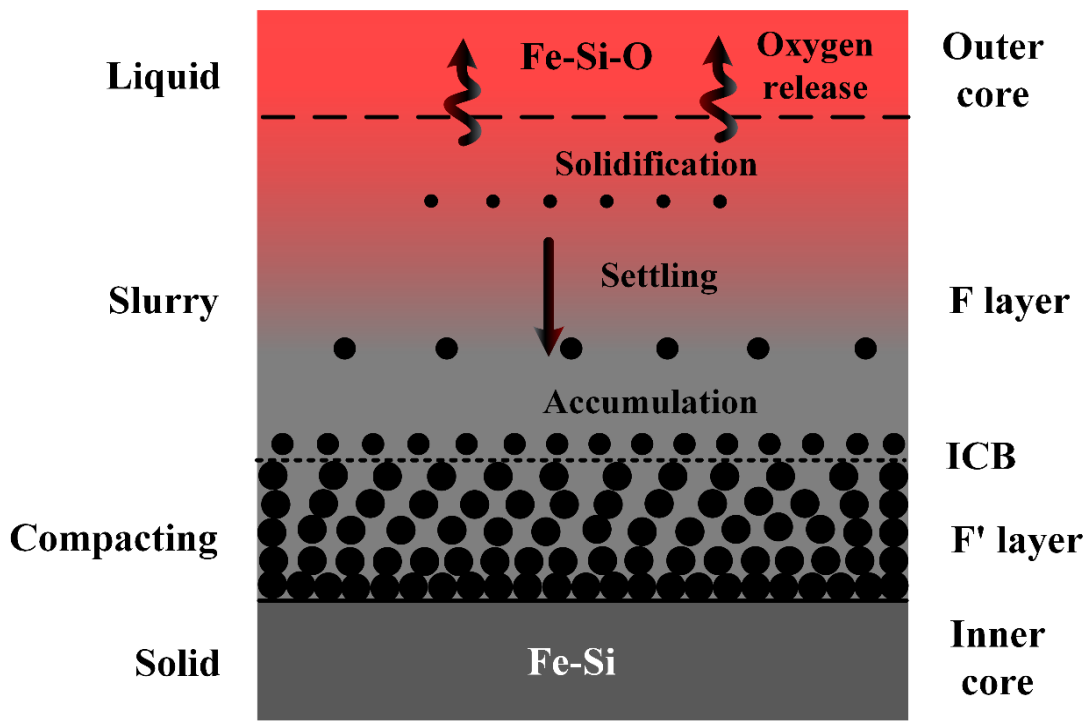

**Fig. 4. Solidification and compaction across the ICB due to the crossover of the core adiabat and the iron alloy liquidus.** Liquid Fe-Si-O in the Earth's outer core crystallizes when its liquidus exceeds the adiabat at the top of the F layer as the core cools. During crystallization, oxygen preferentially partitions into liquid (Alfè et al., 2002; Hirose et al., 2017), forming Fe-Si solid particles. A mixture of solid-liquid

phases exists as a result of crystal suspension in the F layer (slurry). Solid particles fall like snow and accumulate at the bottom of the F layer. A gradient of oxygen concentration is formed due to preferential partitioning, represented by red-grey scale shading where red and grey shadows indicate O distribution from rich to poor. At a radius of 1500 km close to the top of the F layer, the concentration of the light elements is the same as the outer core (~6–8 wt.%). At the bottom of the F layer (1220 km), the composition is close to the inner core's (~3–5 wt.%) as shown in Fig. 3. Compaction occurs in the F' layer as a result of accumulation. The material solidifies completely below the compacting cumulate layer. Solid circles show solid Fe-Si alloy particles.

Sinking crystals are denser than the liquid due to the systematics of light-element partitioning in addition to the volume reduction caused by the phase change. This process results in a stably stratified composition with a light-element concentrated gradient across the F layer (Fig. 3, solid green line). Below the ICB, the adiabat temperature becomes close to the solidus of the inner core, where most of the liquid solidifies and forms a liquid-solid boundary. Considering a ternary Fe-Si-O system as the composition of the Earth's outer core, Fe-Si crystalizes in the F layer so that O partitions from a solid into a liquid, forming a solid-liquid slurry (Fig. 4). The released light element (oxygen) rises into the liquid above the F layer. Solid Fe-Si particles sink down and continuously become a larger and larger proportion of the slurry, which may occur by grain growth and by an increase in the number of crystallizing particles. Eventually, a critical packing fraction is reached where sinking crystals pile up at the ICB, and the particles form a compacting cumulate pile (F' layer) below the F layer. Within the F' layer, the solid matrix continuously compacts under its own weight expelling trapped liquid. Because the F' layer is supported by an interconnected network of solid particles, it exhibits some rigidity. Two important consequences of the compacting would be a density jump (Δ$\rho$) and an increase in P-wave velocity at the boundary between the F and F' layers. The critical packing fraction at the boundary may be 50% to 70% according to studies of crystal accumulation in magma chambers (McKenzie, 2011). The compacting interface is still partially molten and is therefore permeable. The core becomes effectively solid below the base of the compacting pile, which contributes to the inner core's growth.

**3.2. Phase diagram of Fe-Si-O system**

The temperature-light element content phase diagram of Fe-Si-O was investigated from 313 GPa to 329 GPa, corresponding to the radius of 1500 km to 1220 km, respectively, in the F layer (Fig. 5). At the top of the F layer (313 GPa), the temperature is equal to the liquidus temperature of Fe-Si-O alloy; while at the ICB (329 GPa), it is close to the solidus temperature. Here, Fe-4wt.%Si-1.5wt.%O is used for the outer core composition because its $V_p$ displays velocity profiles similar to PREM data in the Earth's outer core (Fig. 7). We molded its solidus and liquidus at the relevant P-T conditions of the ICB based on recent laboratory data (Anzellini et al., 2013; Fischer, 2016; Huang et al., 2010; Morard et al., 2014; Zhang et al., 2018).

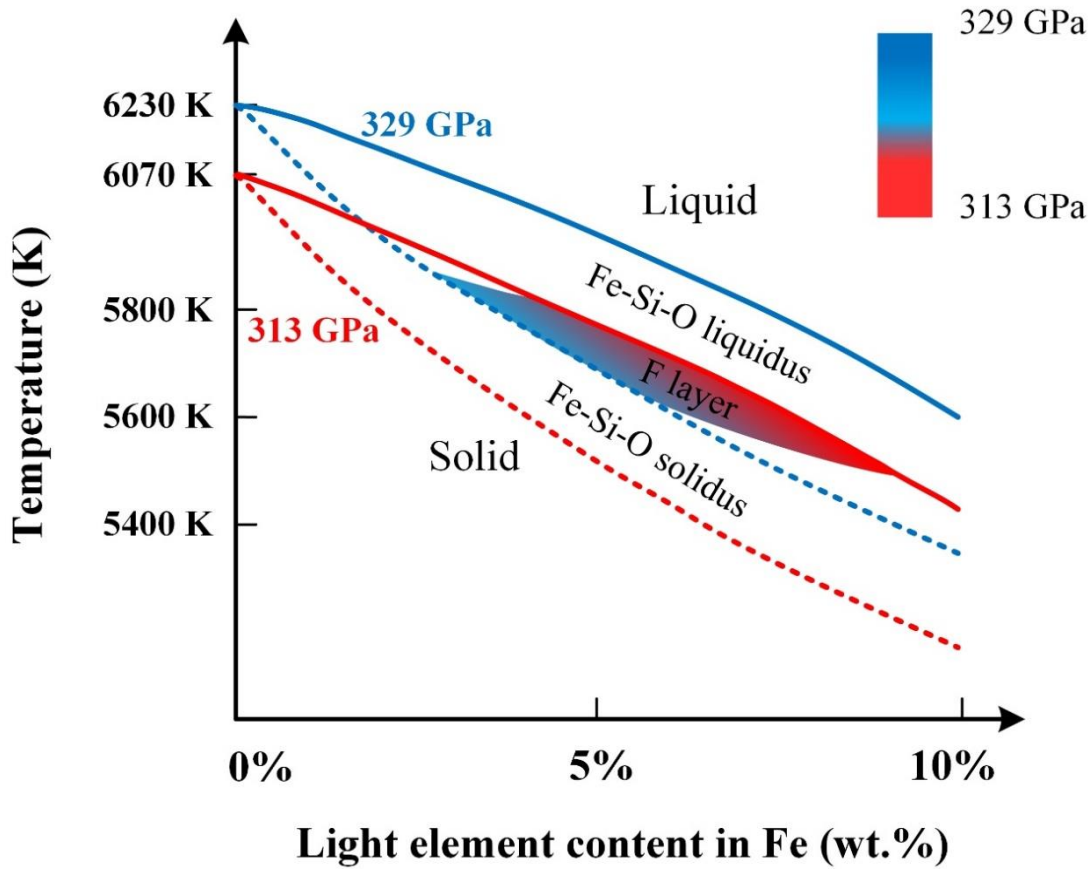

**Fig. 5. Solidus and liquidus of the Fe-Si-O system with varied light element content in Fe at the relevant conditions of the F layer.** The Fe-Si-O liquidus and solidus could be obtained from recently measured melting temperature of Fe, Fe-Si, and Fe-O systems at high pressures (Anzellini et al., 2013; Fischer, 2016; Huang et al., 2010; Morard et al., 2014; Mori et al., 2017; Zhang et al., 2018). From the top (313 GPa) to the bottom (329 GPa) of the F layer, the adiabat temperature moves from Fe-Si-O liquidus close to solidus. The colorful shadow area indicates the change of light element concentration under the conditions of the F layer, where the red gradually changes to blue to represent a pressure gradient from 313 to 330 GPa. Note that the exact widths of the phase loops are not fully determined and will require further refinements in the future. The solid upper-line and dashed down-line represent the liquidus and the solidus as a function of the Si-O concentration at 313 GPa (red lines) and 329 GPa (blue lines), respectively.

At the top of the F layer, the solidus and liquidus for an Fe-Si-O system used in this study (*e.g.*, Fe-4wt.%Si-1.5wt.%O) are estimated to be ~5300 K and ~5650 K from melting temperatures of an O-rich Fe alloy (*e.g.*, Fe-8O-2S system (Huang et al., 2011)) and Si-rich Fe alloy (*e.g.*, Fe-8Ni-10Si system (Zhang et al., 2018)) at ~313 GPa, respectively. Thus, the temperature at top of the F layer is ~5650 K from the modeled liquidus temperature of Fe-4wt.%Si-1.5wt.%O. At the ICB, most of O could be first partitioned into the liquid due to the higher liquidus temperature of Fe-Si, so the temperature at the ICB is ~5800 K determined from melting temperatures of Fe-O alloy (Huang et al., 2011; Komabayashi, 2014). The temperatures within the F layer are between the liquidus of the outer core composition and the solidus of the inner core composition, as shown in the shaded area of Fig. 5. The liquidus gradient of Fe alloy with a compositional gradient and the adiabat gradient in the F layer can be constrained linearly according to the phase diagram (Supplementary Table 1).

### 3.3. Solid fraction in the F layer

We examined the solid fraction in the F layer and compared the compressional sound velocity gradient for the slurry with seismic observations. To model the potential solid fraction in the F layer slurry as a function of radius, we use an equation for energy conservation described by (Gubbins et al., 2008; Malkus, 1972):

$$\frac{df}{dr} = \frac{C_P}{L + P(\rho_p - \rho_f)/\rho_s^2} \frac{dT}{dr} \qquad (1)$$

where $C_p$ is the specific heat, $L$ is the latent heat, $dT$ is the temperature difference between the crystallized solid particles and the adjacent liquid, $df$ is the difference of the liquid fraction, $P$ is the pressure, and $\rho_p$ and $\rho_f$ are the densities of the solid and liquid alloys, which assumes adiabatic conditions. The latent heat of the Fe-alloy crystallization at the ICB conditions is taken from 500–750 kJ/kg (Buffett et al., 1996;

Gubbins et al., 2011; Labrosse, 2003; Nakagawa and Tackley, 2014; Poirier, 1994b). The other parameters used for the calculations are listed in Supplementary Table S1. Within our model, solid particles fractionate from the slurry and solidification produces a compositional gradient in the liquid, deviating from adiabatic conditions. Even so, the simple adiabatic energy balance assumed in Eq. (1) demonstrates the potential for crystal fraction and variation with depth in the F layer slurry.

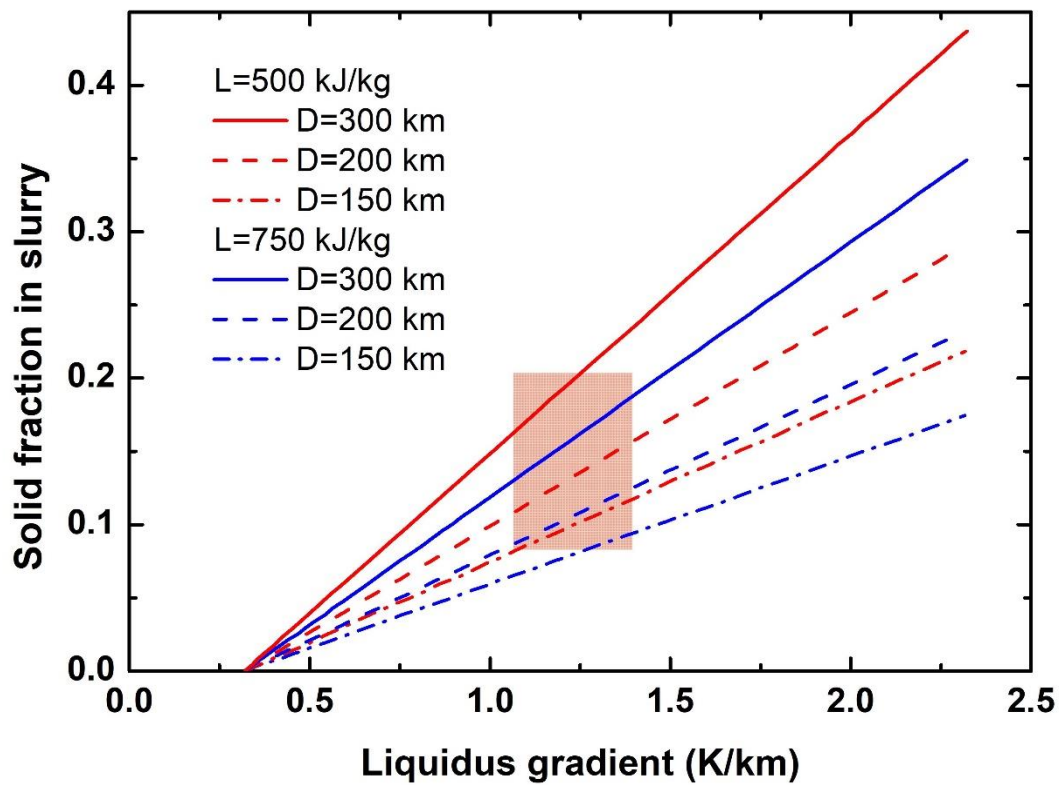

**Fig. 6. Solid fraction in the F layer slurry as a function of liquidus gradient, latent heat ("L"), and the thickness of the F layer ("D").** The solid fraction in the F layer has a positive relationship with the latent heat, the liquidus gradient of Fe alloy, and the thickness of the F layer. Taking a latent heat of 500–750 kJ/kg (Buffett et al., 1996; Gubbins et al., 2011; Labrosse, 2003; Nakagawa and Tackley, 2014; Poirier, 1994b) and a liquidus gradient of ~1.2 K/km for Fe alloy at a ~280 km thickness of the F layer, we obtain a solid fraction of ~15±5% in the F layer slurry. The shadowed area represents the possible solid fraction at the F layer conditions.

The calculated solid fraction as a function of the liquidus gradient, latent heat, and F layer thickness are shown in Fig. 6. The liquidus gradient of Fe alloy with a compositional gradient across the F layer (*e.g.,* Fe-4wt.%Si-1.5wt.%O liquidus towards close to Fe-4wt.%Si liquidus) ($dT_m/dr$) is approximately -1.2 K/km, which is ~2 times higher than that estimated by Gubbins et al. (2008). They did not consider the potential increase in the melting temperature caused by light-element partitioning. For a ~280 km thickness of the F layer, our model gives a solid fraction of about 15±5% in the slurry. The solid fraction is much higher than previous estimates, which predicted a maximum solid fraction of 1–2% in the F layer (Gubbins et al., 2008).

**3.4. Sound velocity of Fe-Si-O slurry in the F layer**

To investigate the reduced compressional-wave velocity in the F layer, the $V_p$ of Fe-Si-O slurry was calculated based on previous experimental and/or modeled data of the densities and compressional sound velocities of *hcp*-Fe (Dewaele et al., 2006; Lin et al., 2005; Mao et al., 2012), liquid Fe (Brown and McQueen, 1986; Ichikawa et al., 2014; Nguyen and Holmes, 2004), *hcp* Fe-9wt.%Si (Fischer et al., 2014), liquid Fe-9wt.%Ni-10wt.%Si (Zhang et al., 2016), and liquid Fe-8wt.%O-2wt.%S (Huang et al., 2011) at the core conditions using the "Reuss average" model. The "Reuss average" model can effectively describe the sound velocity differences between pure liquid and two-phase suspension (slurry), as verified by ultrasonic measurements of the sound velocities for model suspensions and ice slurry (Langlois et al., 2011). The $V_p$ was calculated using the equation:

$$\rho V_p^2 = K + \frac{4G}{3} \tag{2}$$

where $\rho$ is density and $K$ and $G$ are the bulk and shear moduli, respectively. In a slurry or suspension liquid, the shear modulus $G$ approximately equals 0. So, we could obtain:

$$V_{susp} = \sqrt{\frac{K_e}{\rho_e}} \tag{3}$$

where $K_e$ is the effective elastic modulus of the two-phase slurry; $\rho_e$ is the effective density.

$$\rho_e = \rho_p x_p + (1 - \rho_p) x_f \tag{4}$$

Moreover, $K_e$ is obtained using "Reuss average" model here, which assumes uniform stress throughout the medium:

$$\frac{1}{K_e} = \frac{x_p}{K_p} + \frac{(1-x_p)}{K_f} \tag{5}$$

The $V_p$ of Fe-4wt.%Si-1.5wt.%O generally matches the PREM of the outer core with a downwards trend towards the top of the F layer in Fig. 7 (magenta dash line), while it is significantly faster than the AK135 and Zou 2008 models. When the liquid crystalizes and 1.5 wt.% O gradually partitions into the liquid, the solid fraction of Fe-4wt.%Si particles ideally ranges from 0 to 15% towards the F layer bottom. The calculated $V_p$ of the solid-liquid mixture is shown in Fig. 7 (blue dash-dot line), which gradually reduces and matches the AK135, Zou 2008 and Adam 2018 models. Although there are uncertainties both in the mineral physics data and seismic profiles, their overall trends are consistent with each other.

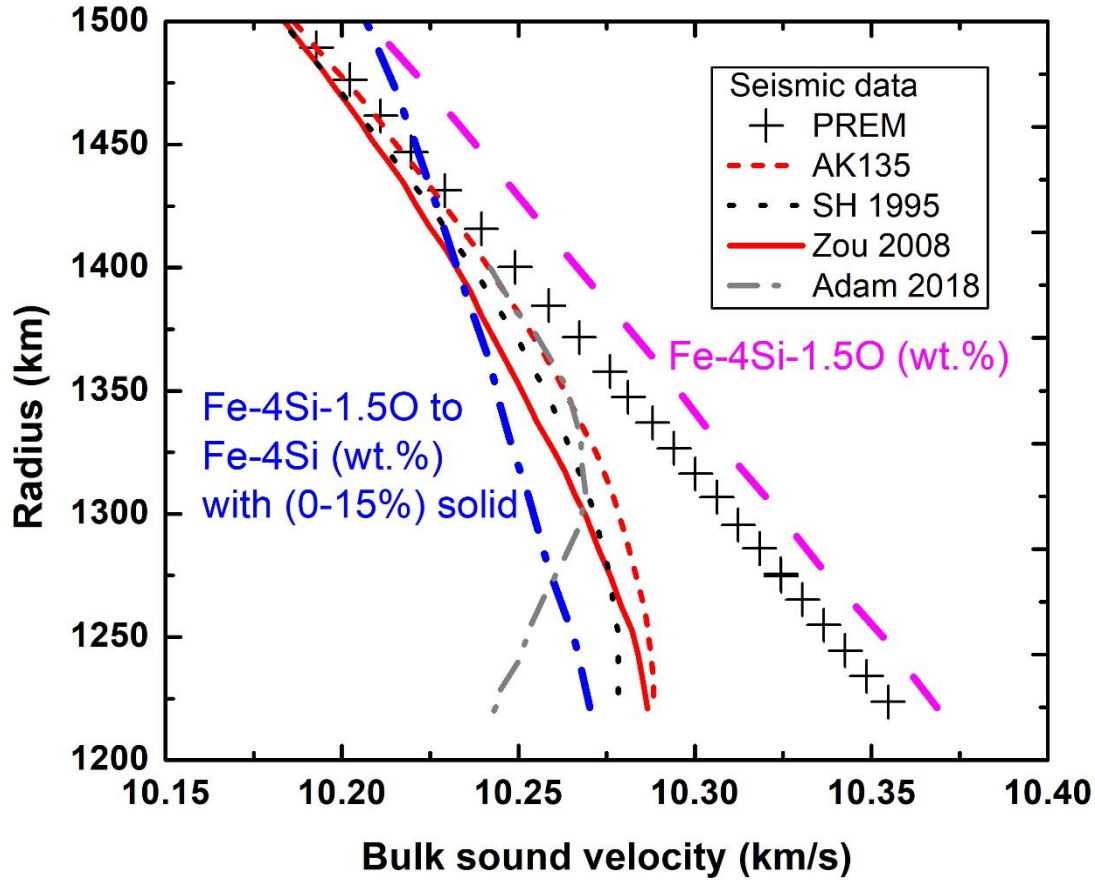

**Fig. 7. Compressional sound velocities of the Fe-Si-O slurry in the F layer compared with seismic data.** The sound velocity of liquid Fe-4wt.%Si-1.5wt.%O generally matches with the PREM data in the outer core but is significantly faster than the AK135, SH 1995, Zou 2008 and Adam 2018 data in the F layer. Considering O partitioning during crystallization and rising into the outer core above the F layer, the composition of Fe-4wt.%Si-1.5wt.%O gradually changes to Fe-4wt.%Si with a maximum of ~15±5% solid fraction towards the bottom of the F layer. The modelled sound velocity of the slurry, mixture of Fe-4wt.%Si-1.5wt.%O liquid and Fe-4wt.%Si solid, could reproduce the Zou 2008 and Adam 2018 data overall.

The compacting cumulate pile that forms the boundary between the inner and outer core occurs when the packing fraction reaches up to ~50–70% due to solid particle accumulation at the bottom of the F layer. The clast-supported framework induces rigidity, transferring the Fe alloy slurry from liquid-like to solid-like (Cates et al., 1998). This transition would produce a sharp increase in seismic velocity of the core as seen at the ICB because of the change in wave-carrying phase from liquid to a phase with non-zero shear modulus. Analogous behavior has been observed in ultrasonic experiments on ice slurry (Langlois et al., 2011) and some dense suspensions (Han et al., 2016).

**3.5. Particle grain size in the F layer**

Our model requires continuous solidification and sinking of crystals in the F-layer. Inner core growth rate places an important constraint on possible accumulation rates and hence the grain size of the Fe alloy snowing in the F layer. We investigated the grain size of solid crystals in the liquid F layer implied by several possible inner-core growth rates (1.22, 0.49, and 0.27 mm/year, corresponding to the inner-core age of 1.0 Gyr, 2.5 Gyr, and 4.5 Gyr, respectively), which were calculated as a function of the viscosity of the liquid outer core using a Stokes's law,

$$r = \sqrt{\frac{9v\eta_f}{2\Delta\rho g}} \qquad (6)$$

where $r$ is the sinking particle radius, $v$ is the effective velocity of the sinking particles (estimated using the inner-core growth rates scaled by the solid fraction in the F layer), $\eta_f$ is the viscosity of the outer core liquid, $g$ is gravitational acceleration (4.40 m/s$^2$ at the ICB) and $\Delta\rho$ is the density difference between solid and liquid. The crystal grain size and liquid outer-core viscosity have a log-linear relationship as shown in Fig. 8. Previous studies argue that the viscosity of the fluid outer core ($\eta_f$) is between 10$^{-2}$ and 10$^2$ Pa·s (de Wijs et al., 1998; Desgranges and Delhommelle, 2007; Palmer and Smylie, 2005; Smylie et al., 2009), corresponding to grain radii between 10$^{-4}$ and 10$^{-2}$ mm, respectively. If we assume the outer core viscosity as ~1 Pa·s and a younger inner core of ~1.0 Gyr recently claimed (Davies et al., 2015), the grain size in the F layer is ~$10^{-3}$ mm at the F layer conditions (Fig. 8). A study of the entrainment of sediments in convecting liquids suggests that if the particle grain size is greater than $3\times10^{-5}$–$10^{-4}$ mm, it would not be entrained by the outer-core thermal convection (Solomatov et al., 1993), indicating that the grain sizes implied by our model would successfully sink to form the inner core for any of the core growth rates we investigated.

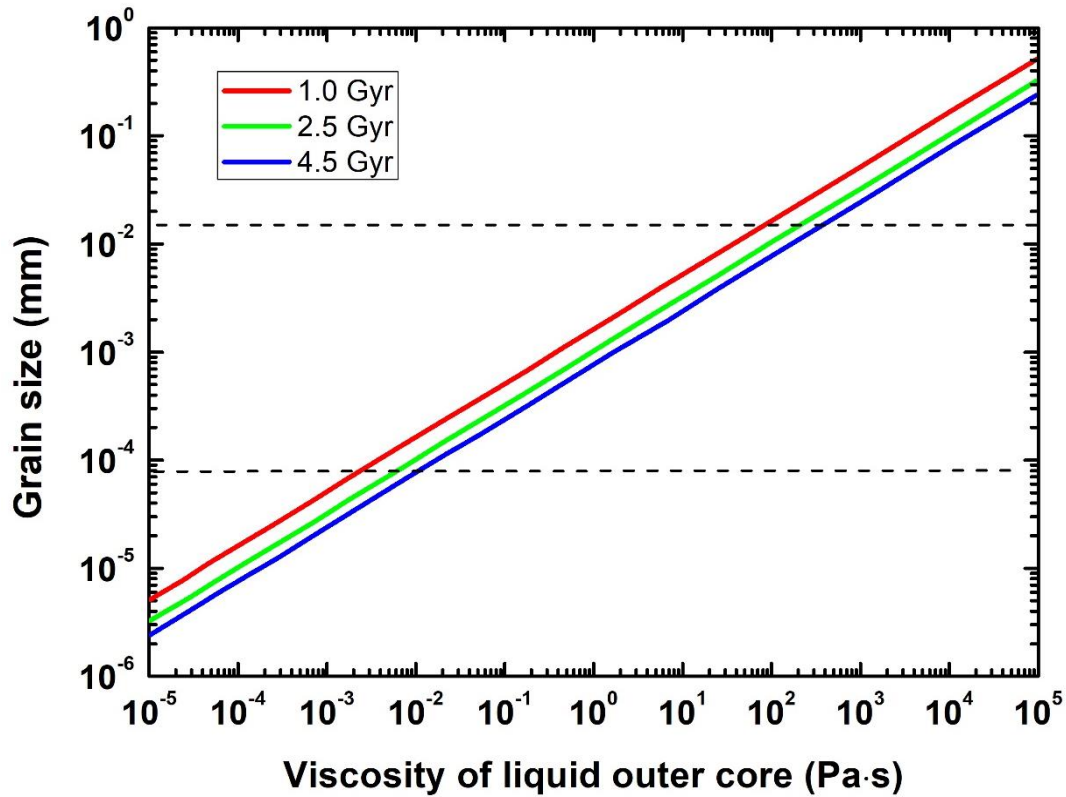

**Fig. 8. The calculated crystal grain size in the F layer as a function of the viscosity in the liquid outer core and inner-core age.** The grain size has a log-linear relationship with the viscosity of liquid outer core. An estimated liquid outer-core viscosity of ~1 Pa·s gives a grain size of ~$10^{-3}$ mm in the slurry F layer. A younger inner core corresponds to larger crystal grain sizes in the F layer. Red, green, and blue solid lines represent the relations between the grain size in the F layer and the viscosity of liquid outer core at the inner-core age of 1.0 Gyr, 2.5 Gyr, and 4.5 Gyr, respectively.

## 4. Discussion

### 4.1. Hemisphere asymmetry in the F' layer

Previous studies show that cold deep subducted lithospheres, such as that beneath Central America (in the western hemisphere) (Van der Hilst et al., 2007), can extract more heat from the core than warm thermochemical piles, such as that beneath the central Pacific (in the eastern hemisphere) (McNamara and Zhong, 2005), indicating a larger heat flux out of the core at the western CMB (Aubert et al., 2008; Van der Hilst et al., 2007). Inhomogeneous conditions at the base of the mantle (D" layer), such as temperature difference (~200–500 K) among the adiabat, subadiabat, and superadiabat (Khan et al., 2008; Nakagawa and Tackley, 2008) as well as anisotropic thermal conductivity of the post-perovskite along different directions in D" region (Ammann et al., 2014), can also produce lateral variation in the CMB heat flux of approximately 10–50% (Ammann et al., 2014; Romanowicz and Gung, 2002; Van der Hilst et al., 2007). As a result, similar heat flux variations on the ICB may be expected because of heat conduction from the inside to out (Gubbins et al., 2011). The temperature variation at the base of the mantle and the heat flux variation across the core is critical in determining the freezing rate of the liquid Fe-Si-O alloy. Therefore, the solidification around the ICB may not be homogeneous (Supplementary material), producing a variation on the thickness of the compacting F' layer in Fig. 9.

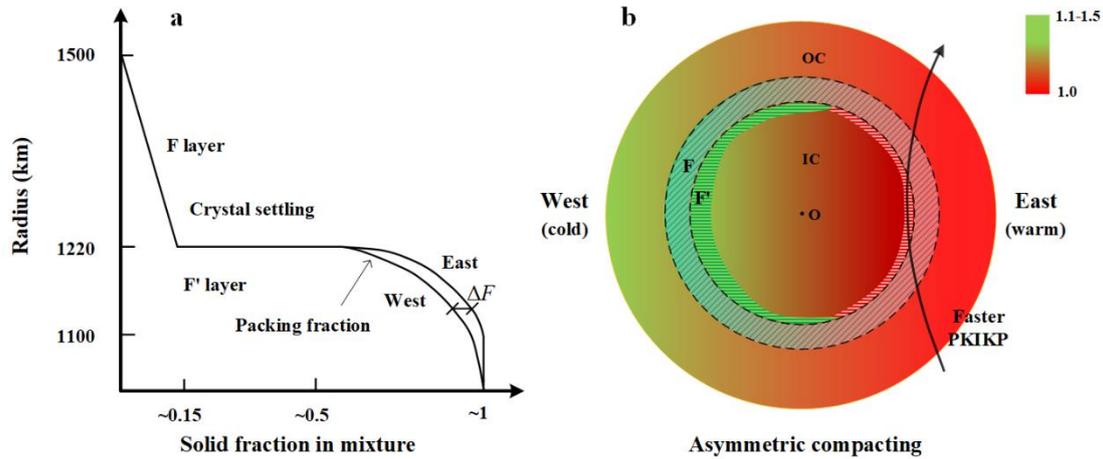

**Fig. 9. Asymmetric solidification and compaction between the eastern and western hemispheres near the inner-core boundary.** The radius versus solid fraction in the Fe alloy solid-liquid mixture phase across the ICB (**a**) and non-dimensional heat flux across the ICB (color contours, where the western hemisphere core in green has approximately 10–50% greater heat flux than that of the eastern hemisphere in red) on the equatorial section of the core (**b**). The east side is warmer than the west side around the base of the mantle (Aubert et al., 2008; Khan et al., 2008; McNamara and Zhong, 2005; Nakagawa and Tackley, 2008; Van der Hilst et al., 2007), which likely causes the hemispheric asymmetry in the thickness of the F' layer. The eastern hemisphere of the inner core has a thinner compacting thickness (~100–150 km) with a faster seismic wave velocity than the western hemisphere (~200–300 km). $\Delta F$ is the solid fraction difference between the two hemispheres in the F' layer.

The solid fraction in the F' layer increases from the critical value (50–70%) to nearly 100% with continuous solidification and compaction (Fig. 9a). Specifically, the relatively higher heat flux in the cold western hemisphere would hasten solidification around the ICB, so the western hemisphere has a higher sedimentation rate (Tkalcic, 2015). Correspondingly, the greater sedimentation rate could result in a thicker compacting pile (Sumita et al., 1996). We, therefore, argue that Earth's western hemisphere has a thicker compacting layer (F' layer) than the eastern hemisphere on average, as shown in Fig. 9b. Consequently, at the same radius in the F' layer, a thin compacting pile in the eastern hemisphere would have a higher solid fraction (Fig. 9a), which could cause a faster sound velocity consistent with seismic observations (Fig. 9b) (Monnereau et al., 2010).

The seismic data show that the P-wave velocity asymmetry between the two hemispheres at the top ~100 km of the F' layer (~1.5%) is much larger than that at its deeper parts (~0.5%) (Fig. 1) (Deuss, 2014; Yu and Wen, 2006). In our model, we propose that it can be explained by variations of the compacting F' layer thickness. At the top ~100 km, the solid fraction difference ($\Delta F$) between the two hemispheres could cause P-wave velocity asymmetry (Fig. 9a). With a mostly solidified eastern hemisphere (such as that below ~150 km of the ICB), the $\Delta F$ would decrease with increasing depth, resulting in the diminution of the P-wave asymmetry between the two hemispheres.

### 4.2. Estimated shear viscosity of the inner core

Interpreting the F' layer as a compacting cumulate pile allows us to estimate the shear viscosity of the solid in the layer. Sumita et al. (1996) conducted a linear analysis to isolate physical controls on the thickness and porosity of the compacting pile. They found that in cases where the rate of particle sedimentation is much less than the Darcy velocity for fluid expelled from the compacting medium (*i.e.*, our F' layer), the layer thickness ($D'$) can be described by:

$$D' = \sqrt{\frac{4\eta_s}{3(\rho_s-\rho_l)g}V_0} \qquad (7)$$

where $\eta_s$ is the shear viscosity of the solid, $g$ is the gravitational constant, $V_0$ is the sedimentation (settling) rate, and $\rho_s$ and $\rho_l$ are the densities of the solid and liquid alloys, respectively. We calculated $V_0$ assuming column geometry, which provided a reasonable estimate for the core growth rate throughout geologic time. A higher $V_0$ exists in the Western hemisphere, which would cause a thicker compacting layer there (Fig. 9).

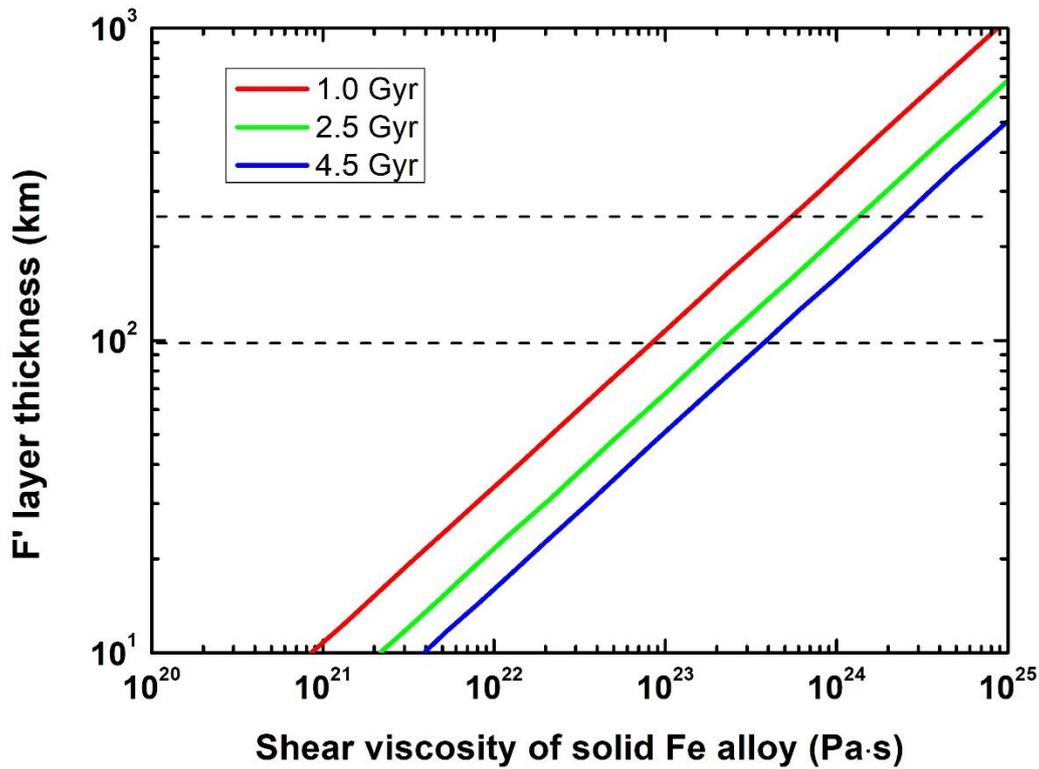

**Fig. 10. Shear viscosity of solid Fe alloy in a quasi-steady state F' layer versus the F' layer thickness and inner-core age.** The calculated shear viscosity of the solid has a log-linear relationship with F' layer thickness. A younger inner core corresponds to a lower shear viscosity of solid Fe alloy. The F' layer has a thickness of ~100–300 km in seismic observations, suggesting a shear viscosity of ~$10^{23}$ Pa·s in the solid inner core. Red, green, and blue solid lines represent the relations between the F' layer thickness and the shear viscosity of solid Fe alloy at the inner-core age of 1.0 Gyr, 2.5 Gyr, and 4.5 Gyr, respectively.

The calculated shear viscosity of the solid Fe alloy in the F' layer is shown in Fig. 10, as a function of the F' layer thickness. For a young inner core of ~1.0 Gyr (inner-core growth rate of ~1.22 cm/year) (Davies et al., 2015), the shear viscosity for the Fe-Si solid in the F' layer is estimated to be ~$10^{23}$ Pa·s according to the observed F' layer thickness of ~100–300 km. This viscosity may be orders of magnitude higher than the aggregate viscosity in shallow parts of the compacting pile, as aggregate viscosity varies with the local liquid fraction (Scott and Kohlstedt, 2006). We infer a ~$10^{23}$ Pa·s shear viscosity for the solid inner core, indicating an insignificant differential rotation of the inner core with respect to the mantle (Buffett, 1997). Thus, the effective viscosity of the F' layer could be located between the viscosities of the fluid outer core ($10^{-2}$–$10^{2}$ Pa·s) and the solid inner core (~$10^{23}$ Pa·s), which is consistent with an derived viscosity of ~$1.22 \times 10^{11}$ Pa·s near Earth's solid inner core through superconducting gravimeter observations (Smylie, 1999).

## 5. Conclusion

In summary, our slurry and compaction model across the ICB provide a simple and reasonable explanation for the observed seismic anomalies and inner-core solidification. This model agrees with recent inferences from mineral physics, geodynamics, and seismic observations. The crystallization of liquid Fe alloy with light element partitioning (*e.g.*, Si and O) can cause a gradually reduced seismic velocity at the lowermost outer core. In addition, lateral variations of the compacting thickness at the upper inner core may induce the west-east asymmetry in P-wave observed by seismology. Our model shows the formation and growth mechanism of the solid inner core and constrains the inner-core shear viscosity.

## Acknowledgments

We thank Bruce A. Buffett for constructive comments that improved this manuscript. Valuable discussions with Jacob S. Jordan at the University of Texas at Austin, T. Sekine and M. Hou at HPSTAR are much appreciated. Thanks to Freyja O'Toole who read our manuscript and made editorial suggestions. Y.Z. acknowledges supports from the National Natural Science Foundation of China (Grant No. 41804082). N.D. acknowledges support of a postdoctoral fellowship provided by the Jackson School of Geosciences, the University of Texas at Austin. J.-F.L acknowledges supports from the Geophysics Program of the National Science Foundation and Deep Carbon Observatory of the Sloan Foundation.

# Supplementary Information for

# Fe alloy slurry and a compacting cumulate pile across Earth's inner-core boundary

**SI Text**

**Parameters in the geodynamic model.** The parameters used in the geodynamic model are shown in Table 1. The density changes were determined from previous experiments and the modelled data for Fe, Fe-Si, and Fe-O alloys (Brown and McQueen, 1986; Dewaele et al., 2006; Fischer et al., 2014; Huang et al., 2011; Ichikawa et al., 2014; Lin et al., 2005; Mao et al., 2012; Zhang et al., 2016) as shown in Fig. 2. The latent heat of Fe-Si alloy has a substantial uncertainty at the core conditions, which could be taken from ~500 to 800 kJ/kg based on previous estimations (Buffett et al., 1996; Gubbins et al., 2011; Labrosse, 2003; Nakagawa and Tackley, 2014; Poirier, 1994). The adiabat gradient in the F layer (~280 km thick) is constrained based on the temperature gap between the top of the F layer and the ICB. The liquidus gradient is contributed by the pressure increase and the light element partitioning in the F layer. The contribution of pressure to the melting is about -0.5 K/km from 313 GPa to 329 GPa (Gubbins et al., 2008), while the contribution from the compositional gradient is about -(0.5 – 1.0) K/km due to light element partitioning. For example, the Fe-4wt.%Si-1.5wt.%O (Fe-4Si-1.5O) and Fe-4wt.%Si (Fe-4Si) are taken as the main compositions at the top and bottom of the F layer, respectively, which correspond to the melting temperatures of ~5650 K and ~6000 K under the related pressures, respectively. Therefore, we could reasonably assume a melting gradient of approximately -1.2 K/km in the ~280 km F-layer.

**The relation between the core-mantle temperature and the inner core growth rate.** The rate of inner core growth $dr_i/dt$ has a simple relationship with the rate of the temperature change $T_c$ at the CMB $dT_c/dt$ (Nimmo, 2015),

$$\frac{dr_i}{dt} = \frac{-1}{\left(\frac{dT_m}{dP} - \frac{dT_a}{dP}\right)\rho_{ICB} g} \frac{T_{ICB}}{T_c} \frac{dT_c}{dt} \tag{S1}$$

where the slopes of the adiabat and melting curve at the ICB are given by $dT_a/dP$ and $dT_m/dP$, respectively, and $T_{ICB}$ and $\rho_{ICB}$ are the ICB temperature and density, respectively. Consequently, the estimated ~200–500 K temperature variation at the current CMB could cause a slight difference in the inner-core growth rate. On the other hand, the latent heat from inner-core solidification and the gravitational energy from light-element release during solidification are the most important sources for CMB heat flow (Lay et al., 2008). Therefore, the variation of the inner core solidification can also affect the $T_c$ at the CMB.

**Residual porosity beneath the compacting cumulate pile.** According to compaction theory, beneath the main part of the compacting cumulate pile (F' layer) a second compaction zone with some residual porosity will be present (Shirley, 1986; Sumita et al., 1996). The residual porosity in this compaction zone is proportional to

$$\sqrt{\frac{\phi_0^3 \eta_f V_0}{K_{\phi 0}(1-\phi_0)\Delta\rho g}} \tag{S2}$$

where $\phi_0$ is the porosity at the critical packing fraction, $\eta_f$ is the viscosity of the fluid, $V_0$ is the settling velocity, $K_{\phi 0}$ is the reference permeability, $\Delta\rho$ is the density difference between solid and liquid and $g$ is gravitational acceleration at the ICB (Sumita et al., 1996). For the grain radii presented in our Stokes' settling analysis in the main text, we calculate reference permeabilities using $\phi_0$ and the permeability law presented by (Wark and Watson, 1998). We obtain residual porosities of lower than 2% for fluid viscosities of the F layer less than $10^2$ Pa·s as shown in Fig. S1. Because this analysis neglects solidification in and beneath the F' layer, this value is an upper bound on residual porosity beneath the F' layer.

**Table S1**. Parameters used for modeling in this study across the inner-core boundary. Our model uses a Si-rich Fe alloy as a composition of the Earth's core, including liquid Fe-4Si-1.5O and solid Fe-4Si as the main constituents of the outer and inner core, respectively.

| Parameters at the F layer | Variable | Value used |
|---|---|---|
| $\Delta\rho$ on melting of Fe alloy without compositional change | $\Delta\rho$ | 0.10–0.15 g/cm$^3$ |
| Pressure/Radius at the ICB | $P_{ICB}$ | 329 GPa/1222 km |
| Pressure/Radius at the top of the F layer | $P_F$ | 313 GPa/1500 km |
| Thickness of the F layer | $D$ | 280 km |
| Pressure gradient in the F layer | $P'$ | -5.71×10$^4$ Pa/m |
| Latent heat of Fe alloy | $L$ | 500–750 kJ/kg* |
| Temperature at the ICB (solidus of Fe-4Si-1.5O) | $T_{ICB}$ | 5800 K |
| Temperature at the top of the F layer (liquidus of Fe-4Si-1.5O) | $T_F$ | 5650 K |
| Liquidus of Fe-4Si at the ICB | $T_{m\text{-}ICB}$ | 6000 K |
| Adiabatic gradient | $dT_a/dr$ | -0.5(0.1) K/km |
| Liquidus gradient of the F layer (Fe-4Si-1.5O liquidus towards Fe-4Si liquidus) | $dT_m/dr$ | -1.2(0.2) K/km |

*Referenced values for the latent heats of Fe-light element alloys are 570 kJ/kg (Poirier, 1994), 600 kJ/kg (Buffett et al., 1996), 625 kJ/kg (Nakagawa and Tackley, 2014), 660 kJ/kg (Labrosse, 2003), and 750 kJ/kg (Gubbins et al., 2011), respectively.

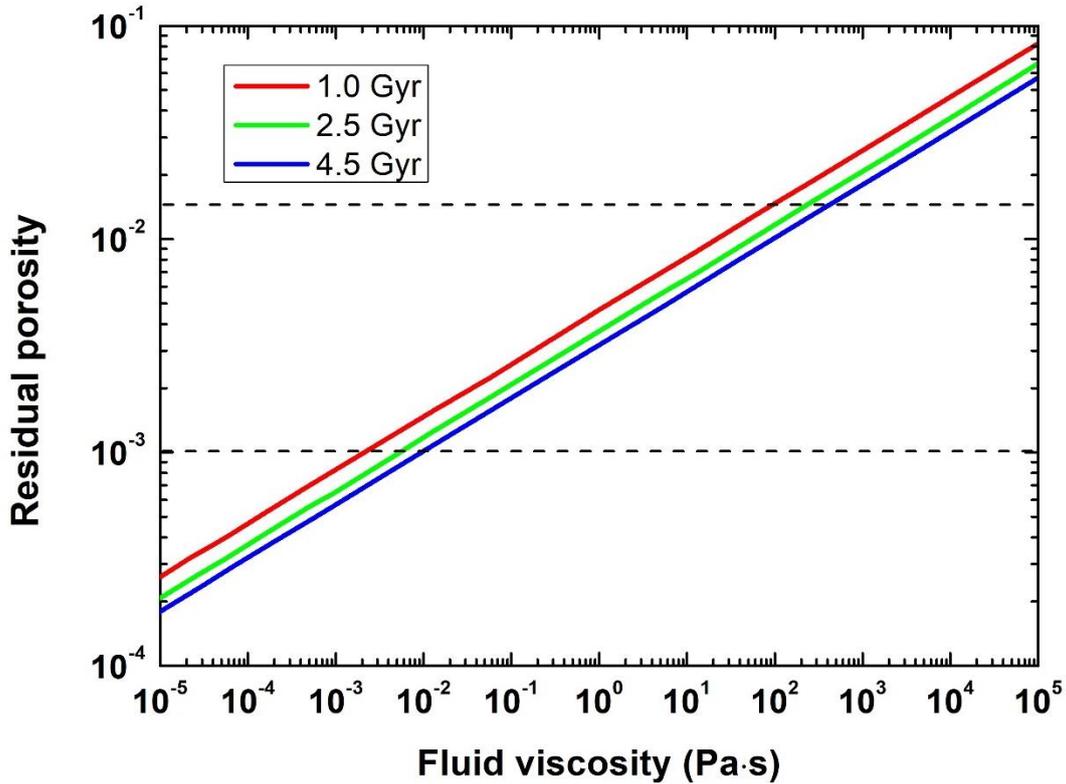

**Fig. S1. Estimated residual porosity beneath the compacting cumulate pile (F' layer) as a function of the fluid viscosity in the slurry layer.** The residual porosities beneath the F' layer is between $10^{-3}$ and $10^{-2}$ for the F layer fluid viscosities from ~$10^{-2}$ to $10^2$ Pa·s. Red, green, and blue solid lines represent the relations between the residual porosity beneath the F' layer and the fluid viscosity at the inner-core age of 1.0 Gyr, 2.5 Gyr, and 4.5 Gyr, respectively.

## SI References